\documentclass[prb,preprint]{revtex4}

\pdfoutput=1

\usepackage{graphicx}

\begin{document}

\title{Origin of quantum oscillations in doped cuprates}

\medskip 

\date{May 6, 2017} \bigskip

\author{Manfred Bucher \\}
\affiliation{\text{\textnormal{Physics Department, California State University,}} \textnormal{Fresno,}
\textnormal{Fresno, California 93740-8031} \\}

\begin{abstract}
It is proposed that Fermi-surface reconstruction in electron-doped $Ln_{2-x}Ce_xCuO_4$ ($Ln = Pr, Nd$) and in hole-doped $YBa_2Cu_3O_{6+y}$ and $YBa_2Cu_4O_8$ occurs when the Fermi arcs extend into the second Brillouin zone (BZ). The criterion employs the axial component of the Fermi-arc tips, $\hat{q} > 0.5$, depending on both the position of the Fermi arc's center $\dot{Q}$ and the incommensurabity $\delta_c$ of unidirectional (striped) charge-density waves (CDWs).
Qualitatively, the concave end-pieces of the Fermi arcs, terminated by Bragg-reflection mirrors due to the CDWs and severed at the boundary of the first BZ by lattice Bragg reflection, are  assumed to join and relax to convex loops. 
Those entities may correspond to the electron pockets attributed to the quantum oscillations observed in these compounds. 
The criterion also explains why \emph{no} quantum oscillations are found in the simple hole-doped lanthanum cuprates,
$La_{2-x}Ae_xCuO_4$ ($Ae = Sr, Ba$),
and in the bismuth cuprates $Bi_2Sr_{2-x}La_xCuO_{6+y}$ and
$Bi_2Sr_2CaCu_2O_{8+y}$. 
The possibility of quantum oscillations in hole-doped, partly substituted $La_{2-y-x}Ln_ySr_xCuO_4$ ($Ln = Nd, Eu; \, y = 0.4, \, 0.2$) in the high-end doping interval of their pseudogap phase, $0.182 < x < 0.235$, is raised.
A geometric modification of Bragg-reflection mirrors applies to $HgBa_2CuO_{4+y}$ where CDWs are bidirectional (checkerboard-like).

\bigskip \bigskip


\end{abstract}

\maketitle

\pagebreak

\section{INTRODUCTION}

Quantum oscillations, as observed by de Haas-van Alphen or Shubnikov-de Haas experiments, have been an invaluable tool in elucidating the Fermi surface of metals.\cite{1} It came as somewhat of a surprise when quantum oscillations were also discovered in the pseudogap phase of the hole-doped cuprates $YBa_2Cu_3O_{6+y}$, $YBa_2Cu_4O_8$, $HgBa_2CuO_{4+y}$, as well as in the electron-doped lanthanide cuprates, $Ln_{2-x}Ce_xCuO_4$ ($Ln = Pr, Nd$).\cite{2,3,4,5,6,7,8,9,10,11,12,13,14,15} The Onsager relation, $F = [\hbar/(2\pi e)]A$, applied to the reciprocal space of their $CuO_2$ planes, relates the quantum-oscillation frequency $F$ to the area $A$ enclosed by the Fermi surface.\cite{1} The observed (relatively) small frequencies $F$ in these compounds have been interpreted as arising from small electron pockets, about 2\% the size of the first Brillouin zone (BZ). 
It is widely recognized that this must result from a Fermi-surface reconstruction. However, no consensus has been reached on the cause of such reconstruction.\cite{7,16,17} Here an explanation is proposed based on Bragg reflection of valence electrons off charge-density waves (CDWs) if the Fermi arc
extends \emph{beyond} the first BZ. 
The electron-doped compounds, $Ln_{2-x}Ce_xCuO_4$, are treated first.
The insights gained will then be used for the hole-doped materials.

\section{ELECTRON-DOPED LANTHANIDE CUPRATES}

A stoichiometric parent crystal $Ln_{2}CuO_{4}$ ($Ln = La, Pr, Nd$) is a Mott insulator with three-dimensional antiferromagnetism (3D-AFM). Doping it with \emph{tetravalent} cerium substitutes ionized lanthanide atoms, $Ln \rightarrow Ln^{3+} + 3e^-$, by ionized cerium, $Ce \rightarrow Ce^{4+} + 4e^-$, causing electron surplus (electron doping) of concentration $n = x$ in $Ln_{2-x}Ce_{x}CuO_4$. Figure 1a shows, as a typical example, the $x$-$T$ phase diagram of $Nd_{2-x}Ce_xCuO_4$.
The doped electrons segregate in the $CuO_2$ plane, partitioning it by pairs of doped electrons and forming an incommensurate superlattice. 
It can be regarded as a static CDW with an incommensurability (being a wave number)
\begin{equation}
\delta_c(x) = \sqrt{\frac{1}{2}}\sqrt{x} \,\,\,\,\,\,\,\,\,\,\,\,\,\,\, (x>x_6)
\end{equation}

\noindent in reciprocal lattice units (r.l.u.).
The formula has been derived in tetragonal approximation of the lattice constants, $a_0 = b_0$, and holds for the doping range $x > x_6 \equiv 2/6^2  \simeq 0.056$ where the CDWs are striped parallel to the crystal's $a$ or $b$ axis.\cite{18} 
Figure 1b shows observed

\pagebreak

\includegraphics[width=5.55in]{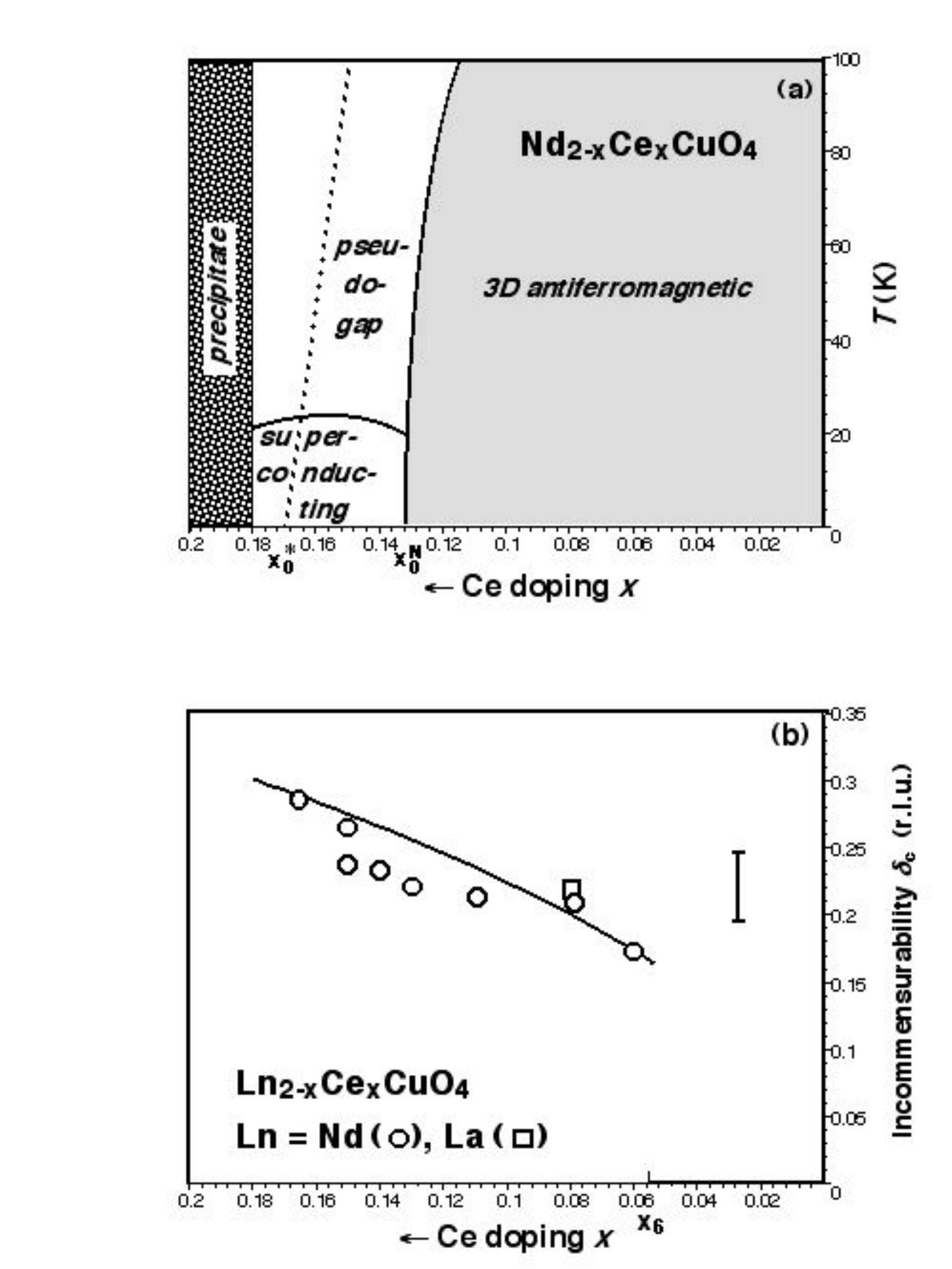}

\noindent FIG. 1. (a) Phase diagram of electron-doped $Nd_{2-x}Ce_{x}CuO_{4}$ (after Ref. 19) with N\'{e}el point $x_0^N$ and quantum critical point $x_0^*$. 
(b) Incommensurability $\delta_c(x)$ of CDWs in $Ln_{2-x}Ce_xCuO_4$ ($Ln = Nd$, crystal, circles; $Ln = La$, film, square) observed with resonant X-ray scattering  (Refs. 20 and 21). The curve is a graph of Eq. (1). An average error bar of the data is shown to the right. 

\pagebreak 
\noindent incommensurabilities of CDWs in ${Nd_{2-x}Ce_{x}CuO_{4}}$ and ${La_{1.92}Ce_{0.08}CuO_4}$, along with a graph of Eq. (1).\cite{20,21} 
In doped $Ln_{2-x}Ce_{x}CuO_{4}$, at a small doping level $x$, a ``metallic'' quantum state (itinerant charge carriers) emerges in reciprocal (``$q$'') space at a certain position $\dot{Q} = (\dot{q},\dot{q})$ on the 2D $q$-space diagonal. It is surrounded by all other quantum states that are still insulating in the 2D-AFM phase on account of an energy gap. For easy reference we want to call $\dot{Q}$ the ``Fermi dot.'' Photoemission spectroscopy (ARPES) and Hubbard-model based calculations show that in $Ln_{2-x}Ce_{x}CuO_{4}$ the (underlying) 2D Fermi surface is hole-like, being centered at symmetry point 
Y = $(\frac{1}{2},\frac{1}{2})$.\cite{22,23,24,25,26,27} In these electron-doped materials the Fermi-dot coordinate has a value of $\dot{q} \approx 0.235 $ r.l.u. with a slight dependence on the lanthanide species $Ln$  but no dependence on $Ce$ doping. The simultaneous occurrence of both metallic and insulating quantum states gives rise to a partial energy gap in the BZ, called ``pseudogap'' (see Fig. 2b).\cite{28}

With increased doping the region of metallic states on the Fermi surface (being a curve in 2D) widens about $\dot{Q}$, leaving a ``Fermi arc'' that bilaterally extends from $\dot{Q}$ out to the arc tips $\hat{Q}$ (see Fig. 2a).\cite{22,23,24,25,26,27} Quantum states along the remaining part of the underlying Fermi surface---from $\hat{Q}$ to the boundary of the BZ---are separated by the pseudogap $\tilde{\Delta} (q, x)$ depending on $q$-space position and doping.
It opens at $\hat{Q}$, where $\tilde{\Delta} (\hat{q}, x) \equiv 0$, and widens in an approximate square-root progression to the boundary of the BZ (see Fig. 2b).

As was noticed early on, at temperatures $T > 0$ the length of the Fermi arc qualitatively increases with doping $x$.\cite{29} For a quantitative assessment it is proposed here that the positions $\hat{Q}$ where the Fermi arc terminates---and the pseudogap opens---are determined by the incommensurability of the CDW, expressed, for $x > x_6$, by the condition for one of the lateral coordinates, say $q = q_a$,
\begin{equation}
\hat{q}(x) = \dot{q} + \delta_{c}(x) \, .  
\end{equation}
Thereby the CDW incommensurability, $\delta_c(x) = \hat{q}(x) - \dot{q}$, provides a rough measure for the length of (each wing of) the Fermi arc for a given doping level $x$ and at $T=0$. As $\delta_{c}(x)$ of Eq. (1) is derived in tetragonal approximation, $a_0 = b_0$, this approximation will be maintained. 

Experiments have shown that CDWs are \emph{unidirectional} in domains of the $CuO_2$ planes of hole-doped $La_{2-x}Ae_xCuO_4$ ($Ae = Sr, Ba$) and
$La_{2-y-x}Ln_ySr_xCuO_4$ ($Ln = Nd, Eu; \, y = 0.4, 0.2$) single crystals of T-type structure.\cite{30} It can be expected that the same holds for the electron-doped compounds $Ln_{2-x}Ce_xCuO_4$ ($Ln = La, Pr, Nd$) of the same `412' family and closely related T'-type structure.\cite{25} For doping $x > x_6$, they are oriented along either the 

\pagebreak
\includegraphics[width=4.32in]{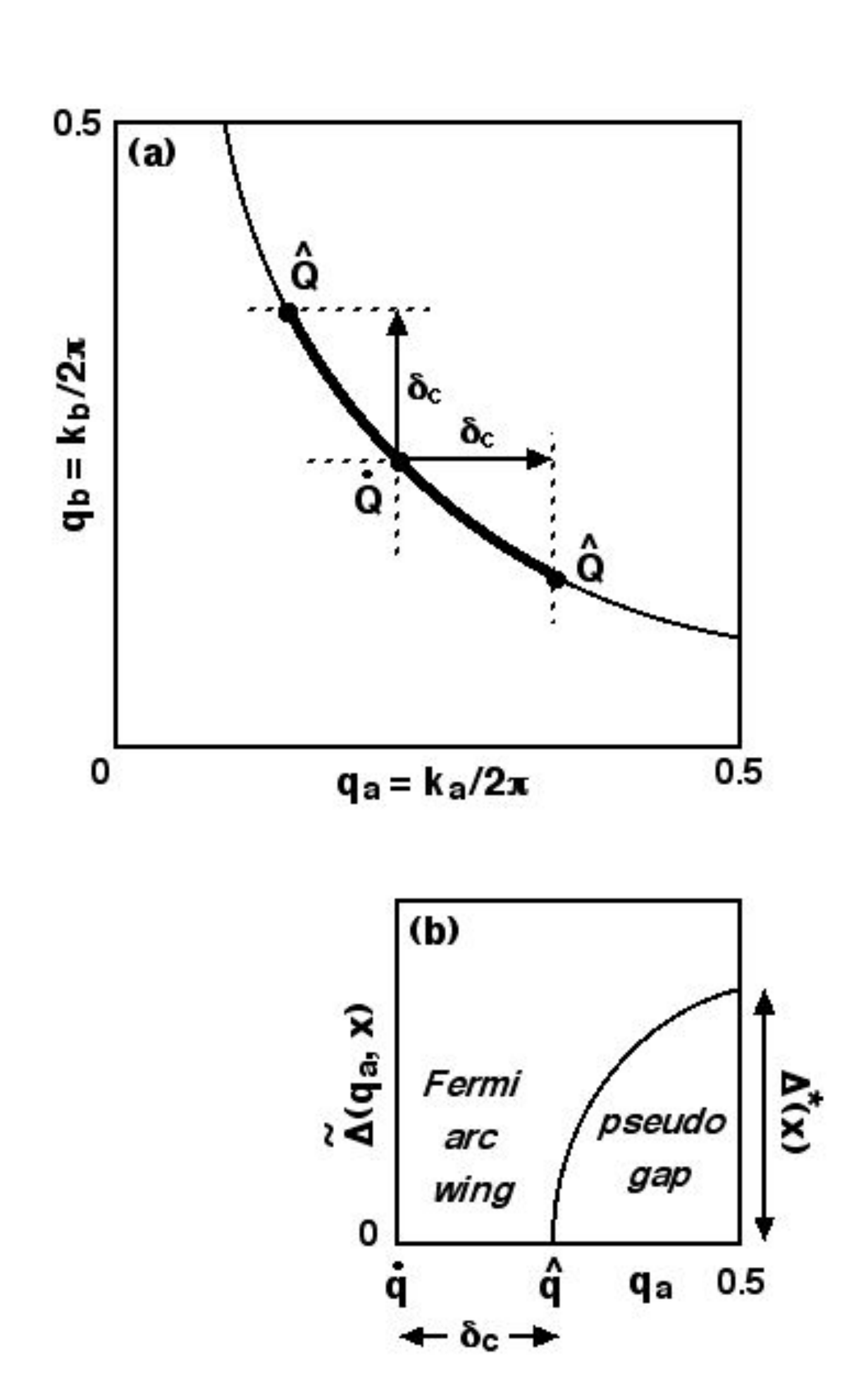}

\noindent FIG. 2. (a) First quadrant of the first Brillouin zone of electron-doped $Ln_{2-x}Ce_{x}CuO_{4}$ ($Ln = La, Pr, Nd$) with (approximate) Fermi arc (bold) and pseudogap states along the underlying (unoccupied) Fermi surface (thin curve). Itinerant holes from quantum states on each wing of the Fermi arc are trapped by unidirectional Bragg-reflection mirrors (dashed) whose extension equals the incommensurability of the CDW, $\delta_{c}(x)$, for a given doping level $x$. (b) Dependence (schematic) of the pseudogap $\tilde{\Delta}(q_a, x)$ on a lateral $q$-space coordinate from the center of the Fermi arc, $\dot{q}$, along a Fermi-arc wing to the wing tip, $\hat{q}$, and further along the underlying Fermi surface to the boundary of the Brillouin zone at $q_a=0.5$.

\pagebreak 

\noindent crystals' $a$ or $b$ direction in respective $a$-domains and $b$-domains. As assumed in the derivation of Eq. (1), the CDW incommensurability can be considered the reciprocal value of the lattice constants $A_0(x) = B_0(x)$ of the doped-hole superlattice, $\delta_c(x)=A_0^{-1}(x)=B_0^{-1}(x)$.\cite{18} This makes $\delta_c(x)$ analogous to the reciprocal value of the crystal lattice constants, $q_{0a} = a_0^{-1}$ and $q_{0b} = b_0^{-1}$. The latter determine the size of the 2D BZ. Likewise $\delta_c(x)$ furnishes the extension of unidirectional CDW  Bragg-reflection mirrors, associated with corresponding $a$-domains and $b$-domains. Attached to $\dot{Q}$, the Bragg-reflection mirrors are spanned by vectors of length $\delta_c(x)$ parallel to the $q_a$ and $q_b$ axes for $x > x_6$ as illustrated in Fig. 2a.  In this way the CDW Bragg-reflection mirrors trap itinerant holes to quantum states on the Fermi-arc wings $\hat{Q}\dot{Q}$ and $\dot{Q}\hat{Q}$.

An important situation is at hand when the tips of the Fermi arc, $\hat{Q}$, touch the boundary of the BZ, 
\begin{equation}
\hat{q}(\overline{x}) \equiv 0.5  \, .
\end{equation}
This is the case at a doping
level 
\begin{equation}
\overline{x} = 2(0.5 - \dot{q})^2  \, ,  
\end{equation}
\noindent obtained by combining Eqs. (1) - (3).
Inserting $\dot{q}$ data from Table I, Eq. (4) gives the $\overline{x}$ values listed in Table I. 

Not only the length of the Fermi arc, but also its \emph{curvature} changes with the doping level $x$ of $Ln_{2-x}Ce_xCuO_4$ as observed with ARPES and confirmed with Hubbard-model calculations.\cite{22,23,24,25,26,27}
Whereas the position of the Fermi dot $\dot{Q} = (\dot{q},\dot{q})$, does not depend on $Ce$ doping and only slightly on the lanthanide species $Ln$ (see Table I), the curvature of the underlying Fermi surface, centered for low and moderate doping at symmetry point Y, relaxes with increased doping.
A result from both increasing arc length and decreasing curvature with more doping, the Fermi arc eventually reaches the antinodal symmetry points M = $(\frac{1}{2},0)$ and $(0,\frac{1}{2})$.
This is the case at the doping level $x_0^*$, marking the high-doping end of the pseudogap phase at $T = 0$, often regarded a \emph{quantum critical point}. Data of $x_0^*$, obtained from experiments with pseudogap probes for $T \rightarrow 0$, as well as of the linear extrapolation from higher-temperature data, $x^*_{0\ell X}$, are listed in Table I. 
The Fermi arc's reaching the M point closes the pseudogap and contiguously joins the Fermi arcs of all quadrants of the first BZ to form a complete Fermi surface with attending metallicity. However, in the present case the metallicity is ``strange'' as explained elsewhere.\cite{31} 

\pagebreak
\includegraphics[width=6.1in]{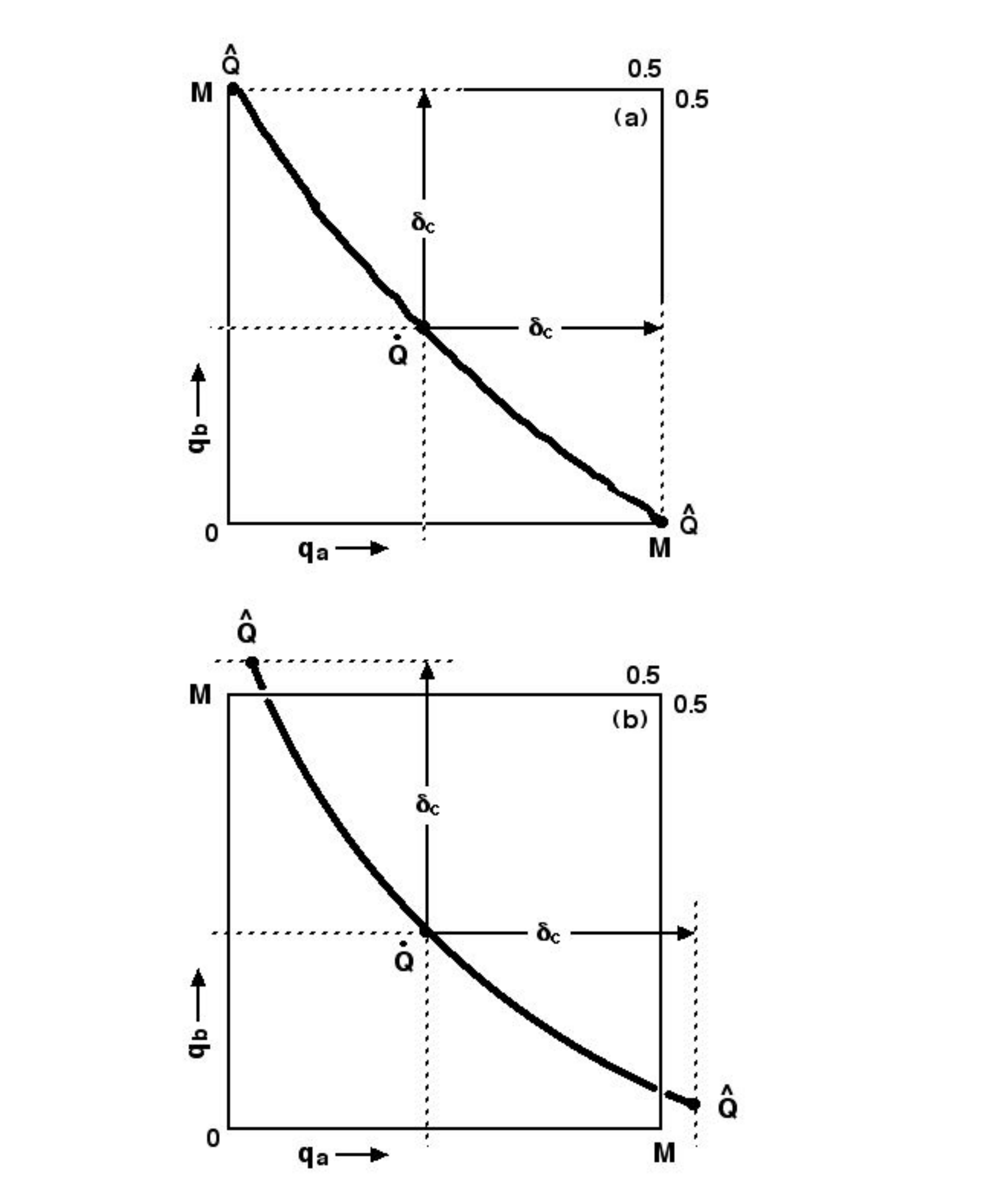}

\noindent FIG. 3. (a) First quadrant of the first Brillouin zone with Fermi arc (schematic) and CDW Bragg-reflection mirrors (dashed) reaching the zone's M points at coinciding doping levels $\overline{x} = x_0^*$ as in $La_{2-x}Ce_{x}CuO_{4}$. 
(b) Fermi arc extending beyond the BZ boundary for doping $\overline{x} < x < x_0^*$ as in $Ln_{2-x}Ce_{x}CuO_{4}$ ($Ln = Pr, Nd$). The pinched-off pieces in the second BZ may reconstruct and combine with equivalent pieces in the other quadrants (not shown) to form ``Fermi ringlets'' that give rise to quantum oscillations. 

\pagebreak

\begin{table}[h!]
\begin{tabular}{ |p{1cm}|p{3.3cm}|p{1.45cm}|p{3.3cm}|p{1.3cm}|p{3.3cm}|p{1.4cm}|  }
 \hline  \hline
 & $La_{2-x}Ce_xCuO_4$ thin film & Ref. & $Pr_{2-x}Ce_xCuO_4$ thin film & Ref. & $Nd_{2-x}Ce_xCuO_4$ single crystal/film & Ref. \\
 \hline  \hline
$x_0^N$ & $0.125 \pm 0.01$ & 32 & $ 0.125 \pm 0.01$ & 33 & $0.135 \pm 0.005$ & 19\\
$\dot{q}$ &   $0.24 \pm 0.01$ & estimate  & $0.24 \pm 0.01$ & 27 & $0.231 \pm 0.005$ & 22,23,26\\
 \hline
$\overline{x}$ & $0.135 \pm 0.01$ & Eq. (4) & $0.135 \pm 0.01$ & Eq. (4)  & $0.145 \pm 0.005$ & Eq. (4)\\
$x_0^*$    & $0.135 \pm 0.01$ & 32 & $0.165 \pm 0.005$ & 34 & $0.175 \pm 0.005$ & 15, 19\\
$x_{0\ell X}^*$ & & & $0.17 \pm 0.01$ & 35, 36 & $0.19 \pm 0.01$ & 23\\
 \hline

$\delta_c(\overline{x})$ & $0.26$ & Eq. (1) & $0.26$ & Eq. (1)  & $0.269$ & Eq. (1)\\
$\hat{q}(\overline{x})$ & $0.50$ & Eq. (3) & $0.50$ & Eq. (3)  & $0.500$ & Eq. (3)\\
$\hat{q}(x_0^*)$ & $0.50 \pm 0.01$ & Eq. (2) & $0.53 \pm 0.01$ & Eq. (2)  & $0.53 \pm 0.005$ & Eq. (2)\\
 \hline   \hline
\end{tabular}
\caption{Electron-doped lanthanide cuprates: N\'{e}el point $x_0^N$ 
(from neutron scattering), Fermi dot $\dot{q}$ (from ARPES), doping level $\overline{x}$ when the Fermi arc touches the BZ boundary, 
quantum critical point $x_0^*$ and linear extrapolation $T \rightarrow 0$ 
of the pseudogap phase boundary to $x_{0\ell X}^*$ (from pseudogap probes), CDW incommensurability $\delta_c(\overline{x})$, Fermi-arc tips $\hat{q}(\overline{x})$ at the BZ boundary, 
and maximal extension of the Fermi arc $\hat{q}(x_0^*)$. 
Reciprocal space coordinates, $q$, and $\delta_c$ are in r.l.u. .}

\label{table:1}
\end{table}

\bigskip  \bigskip

\begin{table}[h!]
\begin{tabular}{ |p{1cm}|p{3.3cm}|p{1.45cm}|p{3.3cm}|p{1.3cm}|p{3.3cm}|p{1.4cm}|  }
 \hline  \hline
$x$ & $La_{2-x}Ce_xCuO_4$ thin film & Ref. & $Pr_{2-x}Ce_xCuO_4$ thin film & Ref. & $Nd_{2-x}Ce_xCuO_4$ single crystal/film & Ref. \\
 \hline
$0.13$ & &  &  &  & $---$ &14\\
$0.14$ & & & $255 \pm 10$ T & 11 & $ $ &\\
$0.15$ & & & & & $290 \pm 10$ T&12, 14\\
$0.16$ & & & & & $285 \pm 15$ T&12, 14\\
$0.165$ & & & & &$270 \pm 10$ T&14\\
$0.17$ & & & & & $250 \pm 10$ T&13, 14\\
$0.175$ & &  &  &  & $---$ extrapolated &15\\
 \hline  \hline
\end{tabular}
\caption{Observed quantum-oscillation frequency $F$ in electron-doped lanthanide cuprates in dependence on the doping level $x$. Note that quantum oscillations occur only in the doping intervals $\overline{x} < x < x_0^*$ from Table I.}

\label{table:2}
\end{table}

\pagebreak

Available data indicate that in the lanthan\emph{um} cuprate $La_{2-x}Ce_xCuO_4$
the doping levels for the Fermi arc's touching the boundary of the first BZ and, respectively, reaching the M points coincide, $\overline{x} = x_0^*$ (see Table I and Fig. 3a). 
However, in the other lanthan\emph{ide} cuprates $Ln_{2-x}Ce_xCuO_4$ ($Ln = Pr, Nd$)
a more curved Fermi arc intercepts, at doping level $\overline{x} \, \, \, (< x_0^*)$, the boundary of the BZ slightly away from (i. e., before) the M points and extends into the second BZ up to its termination by the Bragg-reflection mirror at $\hat{q} > 0.5$ (see Table I and Fig. 3b). Not only are the Fermi-arc segments in the \emph{second} BZ terminated at $\hat{Q}$ by Bragg-reflection mirrors, they are also, due to \emph{lattice} Bragg reflection, pinched off at $q = 0.5$ by the boundary of the lattice BZ---and thus isolated---from the main part of the Fermi arc in the first BZ. 

As a geometric exercise, extend by symmetry the first quadrant of the BZ in Fig. 3b to all four quadrants. Then consider the quadruple of severed Fermi-arc end-pieces in the first and second quadrant near $(0, \frac{1}{2})$ and in the third and fourth quadrant near $(0, -\frac{1}{2})$---and likewise the quadruple of end-pieces in the first and fourth quadrant near $(\frac{1}{2}, 0)$ and in the second and third quadrant near $(-\frac{1}{2}, 0)$.
If the concave pieces of each quadruple are joined, then they form an asteroid-shaped entity.
It is conceivable that such a joining happens in Fermi-arc reconstruction, along with a relaxation and rounding of each asteroid to a convex loop. For ease of reference we want to call the result of this qualitative reconstruction in the \emph{second} BZ ``Fermi ringlets.'' Such Fermi ringlets would provide a necessary condition for \emph{quantum oscillations} in $Ln_{2-x}Ce_xCuO_4$ ($Ln = Pr, Nd$) in the high-end doping interval of their pseudogap phase, $\overline{x} < x < x_0^*$. (The occurrence of quantum oscillations also depends on a sufficiently strong intensity of the CDWs which is not addressed by the present model.)
The Fermi ringlets in the second BZ would be centered at the origin of the BZs, $\Gamma = (0,0)$, and thus be \emph{electron-like}. They might well correspond to the small electron pockets to which the observed small frequencies $F$ of quantum oscillations are attributed.

It is noteworthy that the coincidence of doping levels for BZ boundary-touch and, respectively, M-point-reach, $\overline{x} = x_0^*$, with consequent \emph{lack} of quantum oscillations, in the electron-doped lanthanum cuprate $La_{2-x}Ce_xCuO_4$ is analogous to the situation in the simple hole-doped lanthanum cuprates, $La_{2-x}Ae_xCuO_4$ ($Ae = Sr, Ba$); see Table III.\cite{31} 
On the other hand, the discrimination of doping levels, $\overline{x} < x_0^*$, 
in the other electron-doped lanthanide cuprates, $Ln_{2-x}Ce_xCuO_4$ ($Ln = Pr, Nd$), is analogous to the situation in the hole-doped, partially substituted lanthanum-lanthanide 
cuprates, $La_{2-y-x}Ln_ySr_xCuO_4$ ($Ln = Nd, Eu; \, y = 0.4, 0.2$); see Table III.\cite{31} 
The larger curvature of the Fermi arc in the latter case, $La_{2-y-x}Ln_ySr_xCuO_4$, that makes it reach the boundary of the BZ \emph{before} reaching the M points has been attributed to magnetic contribution from the $Ln^{3+}$ ions' spin magnetic moment $\mu(Ln^{3+})$ instead of $La^{3+}$ ions with none.\cite{31} If this rationale is valid it, then it would equally apply to the electron-doped lanthanide cuprates. 
By the present criterion, quantum oscillations would then \emph{occur} not only in
electron-doped $Ln_{2-x}Ce_xCuO_4$ ($Ln = Pr, Nd$) but also in hole-doped $La_{2-y-x}Ln_ySr_xCuO_4$ ($Ln = Nd, Eu$).
No quantum oscillations have been observed yet in hole-doped $La_{2-y-x}Ln_ySr_xCuO_4$ ($Ln = Nd, Eu; \, y = 0.4, 0.2$). 
The consideration of this possibility is based on the analogy outlined in Table III.

Why are \emph{no} quantum oscillations observed in the simple hole-doped lanthanum cuprates, $La_{2-x}Ae_xCuO_4$ ($Ae = Sr, Ba$)?  
The reason is that in these compounds the Fermi arc never extends into the second BZ, as both contributions to Eq. (2) are too small:
(i) In the hole-doped compounds the observed Fermi-dot position has a lesser value, $\dot{q} = 0.215 \pm 0.005$, than $\dot{q} \approx 0.235$ of the electron-doped materials.\cite{31} (ii) In the hole-doped compounds the CDW incommensurability is given by $\delta_c(x) = \sqrt{\frac{1}{2}}\sqrt{x - x_0^N} \, \, $ ($x>x_6$), instead of Eq. (1), being diminished by the N\'{e}el point $x_0^N = 0.02$ (where 3D-AFM collapses).\cite{18} 
Thus by keeping all of the Fermi arc inside the first BZ, $\hat{q}(x) \leq 0.5$, those hole-doped compounds fail the necessary condition for quantum oscillations.

\bigskip

\begin{table}[h!]
\begin{tabular}{ |p{1.3cm}|p{5.5cm}|p{6.6cm}|p{3.2cm}|  }
 \hline  \hline
 \multicolumn{4}{|c|}{Lanthanide cuprates} \\
 \hline
Doping & Electron-doped & Hole-doped & Comments \\
\hline
$x_0^* = \overline{x}$ & $La_{2-x}Ce_xCuO_4$ & $La_{2-x}Ae_xCuO_4$ ($Ae = Sr, Ba$)  & only lanthanide \\
 & \emph{no} QOs observed & \emph{no} QOs observed &  $La$ involved;\\
 & & &$\mu (La^{3+}) = 0$ \\
 \hline  
 
$x_0^* > \overline{x}$ & $Ln_{2-x}Ce_xCuO_4$ ($Ln = Pr, Nd$) & 
$La_{2-y-x}Ln_ySr_xCuO_4$ ($Ln = Nd, Eu$;  
& other lanthanides \\
 & QOs \emph{observed} for $\overline{x} < x < x_0^*$ & $y = 0.4, \, 0.2$) &  $Ln \ne La$ involved;\\
 & & \emph{possibly} QOs for $\overline{x} < x < x_0^*$ &$\mu (Ln^{3+}) > 0$ \\
 \hline   \hline
\end{tabular}
\caption{Lack and (possible) occurrence of quantum oscillations (QOs) in electron- and hole-doped lanthanide cuprates depending on coincidence or discrimination of doping levels $\overline{x}$ (Fermi arc touching BZ boundary) and $x_0^*$ (quantum critical point). The difference from $Ln \ne La$ may be caused by the difference of spin magnetic moment $\mu$.}
\label{table:3}
\end{table}

\pagebreak

Concluding the electron-doped lanthanide cuprates, $Ln_{2-x}Ce_xCuO_4$ ($Ln = La, Pr, Nd$), the upshot is that quantum oscillations occur in the high-end doping interval of the pseudogap phase, $\overline{x} < x < x_0^*$ (compare Tables I and II),  if the Fermi arc extends into the second BZ. The origin of the quantum oscillation is attributed to reconstructed and joined Fermi-arc segments in the second BZ---called ``Fermi ringlets''---that are severed at the BZ boundary by lattice Bragg reflection from the main Fermi arc in the first BZ.
It can be expected that the area $A$ of a Fermi ringlet, which by Onsager's relation determines the quantum-oscillation frequency $F$, is affected by two counteracting quantities prior to Fermi-arc reconstruction: (i) The extension of the Fermi arc into the second BZ, $\Delta q = \hat{q} - 0.5$, increasing with increased doping $\overline{x} < x < x_0^*$, and (ii) the sideways distance of the arc/boundary intersection from the M point, decreasing with increased doping $\overline{x} < x < x_0^*$. The latter influence seems to be dominant if the observed decrease of $F(x)$ in $Nd_{2-x}Ce_xCuO_4$ is any indication (see Table II). The exact mechanism of the proposed reconstruction of the Fermi-arc end pieces in the second BZ to Fermi ringlets is beyond the scope of the present model. Hubbard-model studies of the Fermi arc may provide more insight.

\section{HOLE-DOPED, OXYGEN-ENRICHED CUPRATES}

Quantum oscillations\cite{2,3,4,5,6,7} and CDWs\cite{37,38,39,40,41,42,43,44,45,46} have also been observed in $YBa_2Cu_3O_{6+y}$. Here the situation is more complicated than in the compounds of the `412' family due to 
\linebreak (i) hole doping through oxygen enrichment $y$ instead of substituting  (infravalent) cation doping $x$ as in $La_{2-x}Ae_xCuO_4$ ($Ae = Sr, Ba$)
and (ii) a crystal structure with \emph{two} $CuO_2$ planes per unit cell and oxygen chains.\cite{30}  
The crystal lattice is tetragonal ($a_0 = b_0$) up to oxygen enrichment 
$y \simeq 0.35$ but orthorhombic ($a_0 < b_0$) beyond.\cite{47}
For about a quarter of the range of oxygen enrichment, 
3D-AFM is stable. After its collapse at $y_0^N \simeq 0.28$, 
magnetic density waves (MDWs) with an incommensurability $\delta_m(y) \propto \sqrt{y - y_0^N}$ emerge---but no CDWs---in the super-oxygenation range $0.28  < y \leq  0.45$.\cite{48}  
The MDWs are observed only along the crystals' $a$ direction (filled triangles in Fig. 4), not along the $b$ direction parallel to the oxygen chains.
At $y = 0.44$, CDWs appear with an incommensurability of $\delta_c^b(0.44) = 0.337$ r.l.u.\cite{45}
The CDWs occur along both the $a$ and $b$ direction, $\delta_c^b(y) > \delta_c^a(y)$ and taper off linearly, by $\sim 10 \%$,
with increasing oxygen enrichment over the range $0.44  \leq y \leq  0.93$ (see Fig. 4).\cite{37,38,39,40,41,42,43,44,45,46} 
To date no analytic expression for $\delta_c(y)$ has been derived.
In contrast to the 

\pagebreak
\includegraphics[width=6.15in]{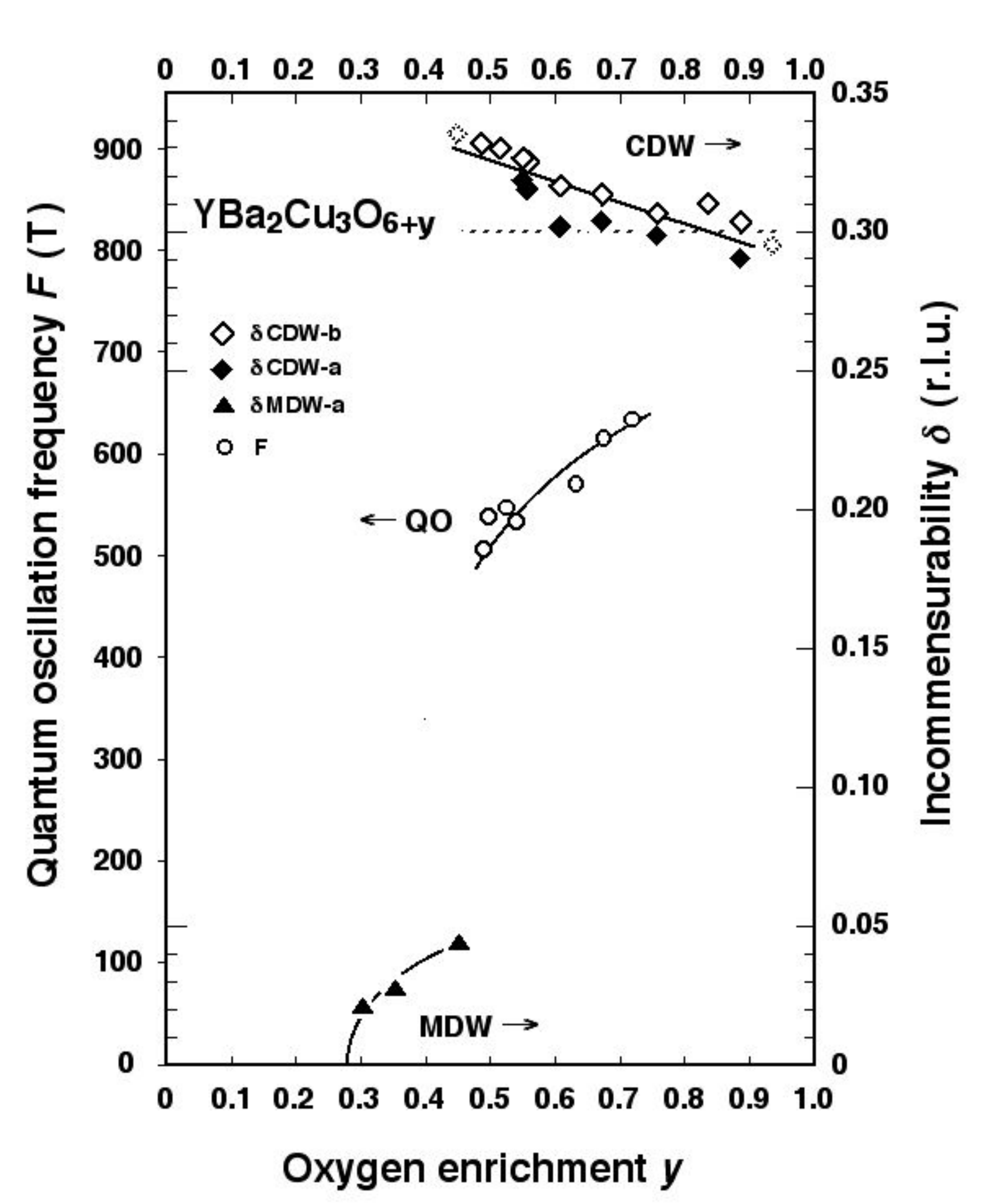}
\footnotesize 
\noindent FIG. 4. Quantum oscillation frequency $F$ (Refs. 2 - 7) and incommensurability $\delta_{m,c}$ of magnetic density waves (MDW) (Ref. 48) and charge density waves (CDW) in $a$ and $b$ crystal directions of $YBa_2Cu_3O_{6+y}$ (Refs. 37 - 46) in dependence of oxygen enrichment $y$. Hatched open diamonds indicate threshold of detection. Solid lines are guides to the eye. If
the average incommensurability
$\delta_c^{ab} \equiv (\delta_c^a + \delta_c^b)/2 > 0.30$ (dashed horizontal line), then the Fermi arc extends into the second BZ, $\hat{q} > 0.5$. This provides, with reconfiguration to Fermi ringlets, a necessary condition for quantum oscillations. 

\pagebreak
\normalsize

\noindent  hole-doped lanthanum cuprates, $La_{2-x}Ae_xCuO_4$ and $La_{2-y-x}Ln_ySr_xCuO_4$ ($Ln = Nd, Eu$), where CDWs and MDWs appear (mostly) together, $\delta_c(x) = 2 \delta_m(x)$, the density waves observed in $YBa_2Cu_3O_{6+y}$ are either one or the other (except for a tiny overlap\cite{44} at $y \simeq 0.445$). The reason for this exclusive appearance of MDWs and CDWs is not known.

The CDW intensity and correlation length exhibit maxima at $y \simeq 0.67$ but fall off in both directions to the limit of detectability (indicated by hatched diamonds in Fig. 4).\cite{44,45} Toward the lower (left) end of super-oxygenation the CDWs along the $b$ direction (open diamonds) are more intensive than those along the $a$ direction (full diamonds). The opposite holds
\bigskip

\begin{table}[h!]
\begin{tabular}{ |p{3.7cm}|p{2.2cm}|p{1.2cm}|p{3.5cm}|p{1.2cm} | }
 \hline   \hline
Compound & $\dot{q}$  (r.l.u.) & Ref. & $\delta_c^{ab}$  (r.l.u.) & Ref.  \\
 \hline
$YBa_2Cu_3O_{6+y}$ & $0.20 \pm 0.01$ & 56 - 58 & $0.32$ --- $0.30 \pm 0.005$ & 37 - 46 \\
$YBa_2Cu_4O_8$ &   $0.20 \pm 0.01$ & 59 & ? & - - - \\
$Bi_2Sr_{2-x}La_xCuO_{6+y}$ & $0.195 \pm 0.01$ & 60, 61 & $0.25 \pm 0.01$ & 16  \\
$Bi_2Sr_2CaCu_2O_{8+y}$ & $0.195 \pm 0.01$ & 62 - 65 & $0.30$ --- $0.25 \pm 0.01$ & 66 \\
$HgBa_2CuO_{4+y}$ & $0.195 \pm 0.01$ &67 - 69& $0.30$ --- $0.26 \pm 0.005$ & 70, 71 \\
 \hline  \hline
\end{tabular}
\caption{Hole-doped, oxygen-enriched cuprates: Fermi dot $\dot{q}$ (from ARPES and Hubbard-model calculations) and average incommensurability of CDWs $\delta_c^{ab} \equiv (\delta_c^a + \delta_c^b)/2$.}

\label{table:4}
\end{table}

\bigskip \bigskip  \bigskip

\begin{table}[h!]
\begin{tabular}{ |p{3.7cm}|p{3.5cm}|p{3.cm}|p{1.cm}|  }
 \hline   \hline
Compound & $\hat{q}$ (r.l.u.) & $F$ (T) & Ref.  \\
 \hline
$YBa_2Cu_3O_{6+y}$ & $0.52 - 0.50 \pm 0.01$ & $505$ --- $636 \pm 20$ & 2 - 7 \\
$YBa_2Cu_4O_8$ &    ? & $660 \pm 30$ & 8, 9 \\
$Bi_2Sr_{2-x}La_xCuO_{6+y}$ & $0.445 \pm 0.01$ & &\\
$Bi_2Sr_2CaCu_2O_{8+y}$ & $0.495 - 0.455 \pm 0.01$ & &\\
$HgBa_2CuO_{4+y}$ & $0.495 - 0.445 \pm 0.01$ & $840 \pm 30$ & 10 \\
 \hline  \hline
\end{tabular}
\caption{Hole-doped, oxygen-enriched cuprates: Fermi-arc tip $\hat{q}= \dot{q} +\delta_c^{ab}$ (from Table IV) and quantum-oscillation frequency $F$.
Note that quantum oscillations occur only if the unreconstructed Fermi-arc tip extends into the second BZ, $\hat{q} > 0.5$. The criterion, based on unidirectional CDW stripes, fails in the case of $HgBa_2CuO_{4+y}$ where CDWs are bidirectional (checkerboard-like).}

\label{table:5}
\end{table}

\pagebreak

\noindent toward the higher (right) end of super-oxygenation with a cross-over at equal intensity at $y \simeq 0.80$.\cite{44}
Quantum oscillations appear at doping $x = 0.49$, at a little more oxygen enrichment than for the first of appearance of CDWs. Their frequency $F$ rises by $\sim 20 \%$ over the range  $0.49  \leq y \leq  0.72$ (see Fig. 4).\cite{7} 
The finding that the smaller super-oxygenation range with quantum oscillations, $0.49  \leq y \leq  0.72$, is embedded in the larger range with CDWs, $0.44  \leq y \leq  0.93$, suggests that the CDWs are too weak on the outer margins of oxygen enrichment to permit quantum oscillations. 
From experiments with pseudogap probes (X-ray diffraction, Kerr effect, Nernst effect, Raman spectroscopy, NMR,  Hall effect, nonlinear optics)\cite{38,39,40,44,45,49,50,51,52,53,54} a quantum critical point (border of the pseudogap phase and strange-metal phase at $T=0$) is extrapolated to oxygen enrichment $y_0^* \simeq 1.00$ (converted from quantum critical hole density per $Cu$ atom in the $CuO_2$ plane, $p_0^* = 0.195 \pm 0.005$).\cite{55} 

We now want to apply the above procedure---Bragg reflection of valence electrons off CDWs---to $YBa_2Cu_3O_{6+y}$. These crystals are not cleavable with a neutral cleavage plane. Accordingly ARPES data are affected by compensating oxygen at the crystal surface. Therefore one has to rely largely on Hubbard-model calculations of the Fermi arc. They provide the Fermi-dot position at $\dot{q} = 0.20 \pm 0.01$ r.l.u.\cite{56,57,58} 
By Eq. (4) we obtain for a super-oxygenation of $y = 0.44$ the Fermi-arc tip at $\hat{q}(0.44) = 0.20 + 0.32 = 0.52 > 0.5$ (see Table IV), extending into the second BZ. 
Again, the terminal Fermi-arc segments are severed at the BZ boundary by \emph{lattice} Bragg reflection (see Fig. 2a).
The severed segments can be expected to reconstruct and join with corresponding pieces in the other quadrants to form \emph{Fermi ringlets} in the second BZ, giving rise to quantum oscillations \emph{if} the intensity of the CDW is strong enough. 
As mentioned, the present model takes into account only the incommensurability, not the intensity of CDWs. Accordingly the presence of Fermi-arc tips in the second BZ, $\hat{q} > 0.5$, is a \emph{necessary}, not a sufficient condition for quantum oscillations.

By Eqs. (2) and (3) the Fermi arc tips touch---from outside and close to the M points---the boundary between the second and first BZ when the average incommensurability, $\delta_c^{ab}(y) \equiv [\delta_c^a(y) + \delta_c^b(y)]/2$,
has decreased to $\delta_c^{ab}(\overline{y}) = 0.5 - \dot{q} = 0.30$, near oxygen enrichment $\overline{y} \simeq 0.80$ (see Fig. 4)---at a somewhat lower super-oxygenation than the observed quantum critical point, $y_0^* \simeq 1.00$.
According to the present Fermi-arc association, the pseudogap would qualitatively close in $YBa_2Cu_3O_{6+y}$ at super-oxygenation $y_0^*$.
The Fermi arc of relaxed curvature then resides completely inside the first BZ, with
Fermi-arc tips joining at the boundary of adjacent quadrants of the first BZ---a topic for future investigations.

\pagebreak

\includegraphics[width=6in]{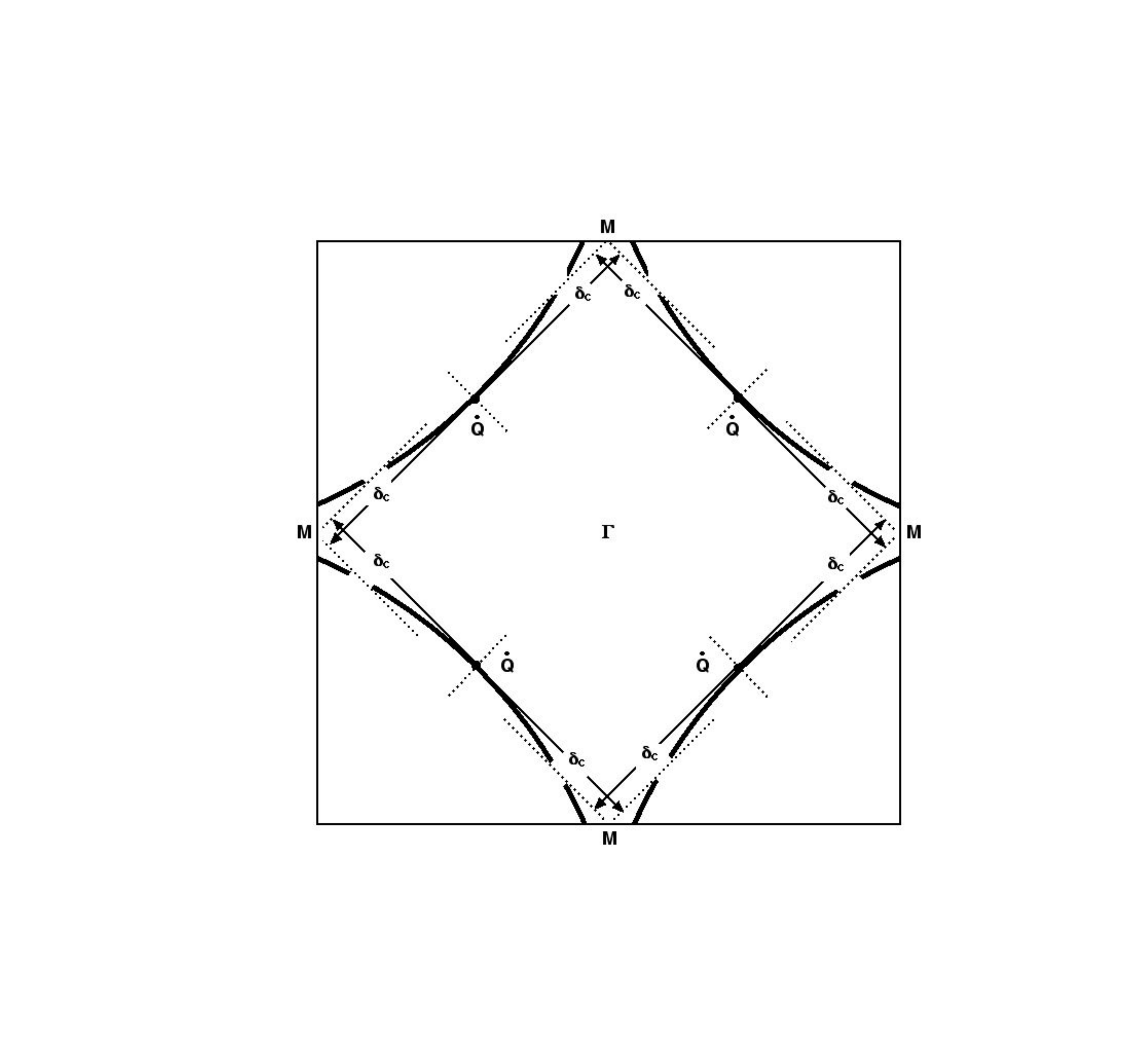}

\noindent FIG. 5. First Brillouin zone with Bragg-reflection mirrors (dotted) spanned from the Fermi dots $\dot{Q}$ by \emph{diagonal} incommensurabilty vectors $\delta_c$. The severed end-pieces of the Fermi arcs may relax, join and reconstruct to Fermi ringlets, providing a necessary condition for quantum oscillations.

\pagebreak

Concerning $YBa_2Cu_4O_8$, a compound with three $CuO_2$ planes per unit cell, quantum oscillations of frequency $F = 660 \pm 30$ T have been observed.\cite{8,9} But \emph{no} detection of CDWs by X-ray diffraction has been reported to date (see Table IV). If the present model is valid, then an incommensurability $\delta_c > 0.5 - \dot{q} = 0.30$ would be expected.

The opposite holds for the bismuth-based cuprates $Bi_2Sr_{2-x}La_xCuO_{6+y}$ and $Bi_2Sr_2CaCu_2O_{8+y}$
where Fermi-dot positions $\dot{q}$ and CDW incommensurabilites $\delta_c$ are both observed, but \emph{no} quantum oscillations (see Tables IV and V). 
This finding is in agreement with the present criterion:
By Eq. (2) the Fermi-arc tips, $\hat{q} < 0.5$, reside within  the first BZ, failing to meet the necessary condition for quantum oscillations.

With a tetragonal lattice and only \emph{one} $CuO_2$ plane per unit cell
$HgBa_2CuO_{4+y}$ has the simplest crystal structure of all superconducting cuprates. Because of its structural simplicity the material is often regarded a model system to elucidate the pseudogap phase. A Fermi arc, CDWs, and quantum oscillations all are observed in $HgBa_2CuO_{4+y}$ (see Tables IV and V). By Eq. (2) the Fermi-arc tips, $0.445 < \hat{q} < 0.495 < 0.5$, reside in the first BZ---in \emph{violation} of the present model's necessary condition for quantum oscillations, $\hat{q} > 0.5$. What could be the reason for this failure?  
As the data of Fermi dot $\dot{q}$ and incommensurability $\delta_c$ 
are reliable, the failure must rest with the assumptions underlying Eq. (2), here unidirectional density waves. The likelihood of \emph{bidirectional} CDW order in $HgBa_2CuO_{4+y}$, with accompanying checkerboard pattern, instead of unidirectional charge-stripe order\cite{30,73,74}
as in $La_{2-x}Ae_xCuO_4$, $YBa_2Cu_3O_{6+y}$ and $Bi_2Sr_{2-x}La_xCuO_{6+y}$, has been raised in Refs. 71 and 72.
In this case the Bragg-reflection mirrors, spanned by \emph{diagonal} vectors $\mathbf{\delta_c} = (\pm \delta_c^a, \pm \delta_c^b)$ and attached to the Fermi dot $\dot{Q}$, intersect the Fermi-arcs \emph{inside} the first BZ (see Fig. 5).
Again, it can be expected that the severed end-pieces of the Fermi arc join, relax and reconstruct to Fermi ringlets. Details of the reconstruction remain subject of future investigations. 

\bigskip \bigskip 

\centerline{ \textbf{ACKNOWLEDGMENTS}}

\noindent I thank Duane Siemens for valuable discussions and Preston Jones for help with LaTeX. 
\linebreak I also thank Neven Bari\v{s}i\'{c} 
and Martin Greven for literature references.


\end{document}